\documentclass[twocolumn, trackchanges]{aastex63}

\graphicspath{{./}{figures/}}

\usepackage{makecell, mathtools, multirow}
\usepackage{hyperref}

\submitjournal{ApJ}

\shorttitle{IFU Spectroscopy of TDE Host Galaxies}
\shortauthors{Hammerstein et al.}

\begin{document}

\newcommand{\galfit}{\texttt{GALFIT}}
\newcommand{\gist}{\texttt{GIST}}
\newcommand{\ppxf}{\texttt{ppxf}}
\newcommand{\msigma}{$M_{\rm BH} - \sigma$}
\title{Integral Field Spectroscopy of 13 Tidal Disruption Event Hosts from the ZTF Survey}

\correspondingauthor{Erica Hammerstein}
\email{ekhammer@astro.umd.edu}
\author[0000-0002-5698-8703]{Erica Hammerstein}
\affil{Department of Astronomy, University of Maryland, College Park, MD 20742, USA}
\affil{Astrophysics Science Division, NASA Goddard Space Flight Center, 8800 Greenbelt Rd, Greenbelt, MD 20771, USA}
\affil{Center for Research and Exploration in Space Science and Technology, NASA/GSFC, Greenbelt, MD 20771, USA}

\author[0000-0003-1673-970X]{S. Bradley Cenko}
\affil{Astrophysics Science Division, NASA Goddard Space Flight Center, 8800 Greenbelt Rd, Greenbelt, MD 20771, USA}
\affil{Joint Space-Science Institute, University of Maryland, College Park, MD 20742 USA}

\author[0000-0003-3703-5154]{Suvi Gezari}
\affiliation{Space Telescope Science Institute, 3700 San Martin Drive, Baltimore, MD 21218, USA}
\affiliation{Department of Physics and Astronomy, Johns Hopkins University, 3400 N. Charles St., Baltimore, MD 21218, USA}

\author[0000-0002-3158-6820]{Sylvain Veilleux}
\affil{Department of Astronomy, University of Maryland, College Park, MD 20742, USA}
\affil{Joint Space-Science Institute, University of Maryland, College Park, MD 20742 USA}

\author[0000-0002-9700-0036]{Brendan O'Connor}
\affil{Department of Physics, The George Washington University, Washington, DC 20052, USA}
\affil{Astronomy, Physics and Statistics Institute of Sciences (APSIS), The George Washington University, Washington, DC 20052, USA}
\affil{Department of Astronomy, University of Maryland, College Park, MD 20742, USA}
\affil{Astrophysics Science Division, NASA Goddard Space Flight Center, 8800 Greenbelt Rd, Greenbelt, MD 20771, USA}

\author[0000-0002-3859-8074]{Sjoert van Velzen}
\affil{Leiden Observatory, Leiden University, Postbus 9513, 2300 RA, Leiden, The Netherlands}

\author[0000-0002-4557-6682]{Charlotte Ward}
\affil{Department of Astrophysical Sciences, Princeton University, Princeton, NJ 08544, USA}

\author[0000-0001-6747-8509]{Yuhan Yao}
\affil{Cahill Center for Astrophysics, California Institute of Technology, MC 249-17, 1200 E California Boulevard, Pasadena, CA, 91125, USA}

\author[0000-0002-3168-0139]{Matthew Graham}
\affil{Department of Astronomy, California Institute of Technology, 1200 E. California Blvd, Pasadena, CA, 91125, USA}

\begin{abstract}
The host galaxies of tidal disruption events (TDEs) have been shown to possess peculiar properties, including high central light concentrations, unusual star-formation histories, and ``green'' colors.
The ubiquity of these large-scale galaxy characteristics among TDE host populations suggests they may serve to boost the TDE rate in such galaxies by influencing the nuclear stellar dynamics.
We present the first population study of integral field spectroscopy for thirteen TDE host galaxies across all spectral classes and X-ray brightnesses with the purpose of investigating their large-scale properties.
We derive the black hole masses via stellar kinematics (i.e., the $M-\sigma$ relation) and find masses in the range $5.0 \lesssim \log(M_{\rm BH}/M_\odot) \lesssim 8.0$, with a distribution dominated by black holes with $M_{\rm BH} \sim 10^6 M_\odot$. We find one object with $M_{\rm BH} \gtrsim 10^8 M_\odot$, above the ``Hills mass'', which if the disrupted star was of solar type, allows a lower limit of $a \gtrsim 0.16$ to be placed on its spin, lending further support to the proposed connection between featureless TDEs and jetted TDEs.
We also explore the level of rotational support in the TDE hosts, quantified by $(V/\sigma)_e$, a parameter which has been shown to correlate with stellar age and may explain the peculiar host galaxy preferences of TDEs.
We find that the TDE hosts exhibit a broad range in $(V/\sigma)_e$ following a similar distribution as E+A galaxies, which have been shown to be overrepresented among TDE host populations.

\end{abstract}

% \keywords{}

\section{Introduction} \label{sec:intro}
It is generally accepted that most, if not all, massive galaxies host a supermassive black hole (SMBH) in their nucleus which play important roles in the evolution and properties of their host galaxies \citep[e.g.][]{Kormendy95, Magorrian99, ho08, gultekin09, kormendy13, Veilleux2005, Fabian2012, Veilleux2020}. This is evident from scaling relations between the SMBH mass and host galaxy properties such as the bulge velocity dispersion \citep[e.g.][]{Ferrarese00, Gebhardt00} or bulge luminosity \citep[e.g.][]{Dressler89, Magorrian98}. These objects can announce their presence most prominently through sustained accretion of nuclear gas and dust as active galactic nuclei (AGN), but many more SMBHs lie dormant, making the study of these objects more difficult. The tidal disruption of a star by the central SMBH, known as a tidal disruption event (TDE), provides a unique way to gain insights on the population of distant and mostly quiescent SMBHs.

A TDE occurs when a star passes sufficiently close (i.e., within the tidal radius) to a SMBH such that the tidal forces felt by the star are stronger than its own self-gravity, resulting in the star being torn apart and roughly half of that stellar debris being eventually accreted by the black hole, creating a luminous flare of radiation potentially visible from Earth \citep{Rees88, evans89,ulmer99}. TDEs were only a theoretical prediction just $\sim$50 yrs ago \citep{Hills75,Lidskii1979}, and we now have observational evidence of these events from the radio to X-rays, with the largest samples of TDEs discovered in the optical using surveys such as iPTF \citep{Blagorodnova2017, Blagorodnova2019, Hung2017}, ASAS-SN \citep{Holoien2014_14ae, Holoien16_14li, Holoien2016b_15oi, Holoien2019_19bt, Wevers2019, Hinkle2021}, Pan-STARRS \citep{Gezari2012, Chornock2014, Holoien2019b_PS18kh, nicholl19}, SDSS \citep{vanvelzen11}, and ZTF \citep{vanVelzen2019_NedStark, vanVelzen21, Hammerstein23, Yao2023}. While the light curves and spectra of TDEs offer important clues to the formation of the accretion disk, winds, and jets, the host galaxies of these transients provide insights into SMBH--galaxy co-evolution, galaxy evolution and mergers, and the dynamics of galaxy nuclei. Understanding the environments that are most likely to host TDEs will even lead to more efficient discovery and follow-up during the era of the Vera Rubin Observatory, which is predicted to observe hundreds to even thousands of new TDEs a year \citep{vanvelzen11}.

TDEs have also been shown to be observed preferentially in E+A or post-starburst galaxies \citep{Arcavi14, French16, LawSmith17, Hammerstein2021}, whose optical spectra are characterized by little to no H$\alpha$ or [\ion{O}{2}] emission and strong Balmer absorption, indicating the presence of stars formed within the past Gyr but no current star formation activity. Typical E+A overrepresentation (i.e., the ratio between the fraction of TDE hosts that are E+As to the fraction of all galaxies that are E+As) ranges widely depending on the study, with some population studies finding an overrepresentation of over 100$\times$ \citep{LawSmith17} and others finding an overrepresentation of just 22$\times$ \citep{Hammerstein2021}. E+A galaxies are also known to have large bulge-to-light ratios, high S\'ersic indices, and high concentration indices \citep{Yang08}, all of which have been shown to greatly enhance the TDE rate in these galaxies by making more stars available in the nuclear region to be tidally disrupted \citep{Stone16a, stone16, French20}.

Several previous studies have aimed to characterize the environments that are most likely to host TDEs and have shown that certain large-scale galaxy properties are indeed linked with higher TDE rates. \citet{graur18} found that TDE host galaxies have higher stellar mass surface density and lower velocity dispersions as compared to a sample of galaxies not known to host recent TDEs. \citet{LawSmith17} examined a sample of TDE host galaxies in comparison to the local galaxy population and found that all of the TDE hosts in their sample reside below the star formation main sequence, have bluer bulge colors, high S\'ersic indices and high bulge-to-light ratios compared to galaxies of similar masses. \citet{Hammerstein2021} found that 61\% of TDE host galaxies in their sample were in the green valley between the star-forming ``blue cloud'' and the passive ``red sequence'' of galaxies, compared to only 13\% of SDSS galaxies. They also found that while most green valley galaxies have S\'ersic indices comparable to blue cloud galaxies, the TDE hosts had higher S\'ersic indices most similar to red, passive galaxies. All of these properties are indicative of systems which have undergone a merger that produce concentrated central stellar distributions and can indeed enhance the TDE rate \citep{Stone16a, stone16, French20}.

In this paper, we present integral field spectroscopy (IFS) of a sample of 13 TDE host galaxies from the Zwicky Transient Facility (ZTF) survey in order to obtain their black hole masses and understand their large-scale kinematics and stellar populations, the latter of which we compare to several other galaxy populations, including E+A galaxies. Integral field spectroscopy provides spatially resolved spectra which gives a study such as this one an edge over long-slit spectroscopy when attempting to probe various size scales of the TDE host galaxies. In Section \ref{sec:observations}, we describe the observations of the 13 TDEs in our sample as well as the subsequent data reduction and analysis methods. We present the results of the kinematic and stellar population analysis and discuss these results in Section \ref{sec:results}. We discuss the results pertaining to the black hole mass in Section \ref{sec:bhmass} and those pertaining to the stellar kinematics and populations in Section \ref{sec:kinematic}. We close with our conclusions in Section \ref{sec:conclusions}.

\section{Observations \& Data Analysis} \label{sec:observations}
\begin{figure*}
    \centering
    \includegraphics[width=\textwidth]{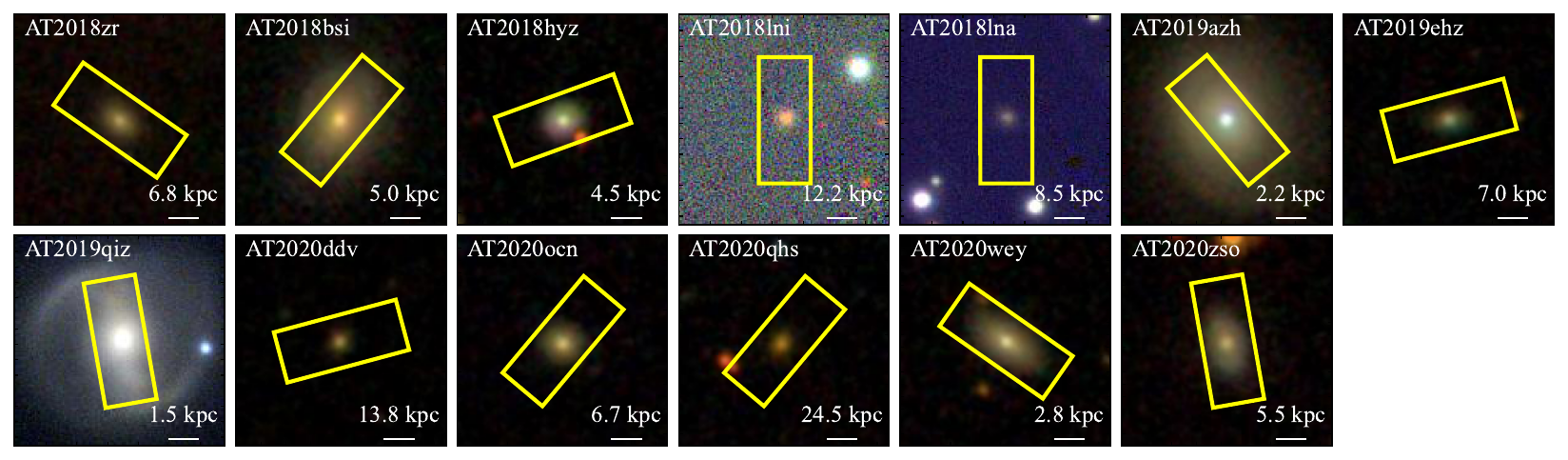}
    \caption{SDSS and Pan-STARRS \textit{gri} images of the thirteen TDE host galaxies, with the yellow rectangle representing the positioning of the KCWI field of view. All images are $34\arcsec \times 34\arcsec$ and the KCWI field of view is $8\farcs4 \times 20\farcs4$.}
    \label{fig:hosts}
\end{figure*}

\begin{figure}
    \centering
    \includegraphics[width=\columnwidth]{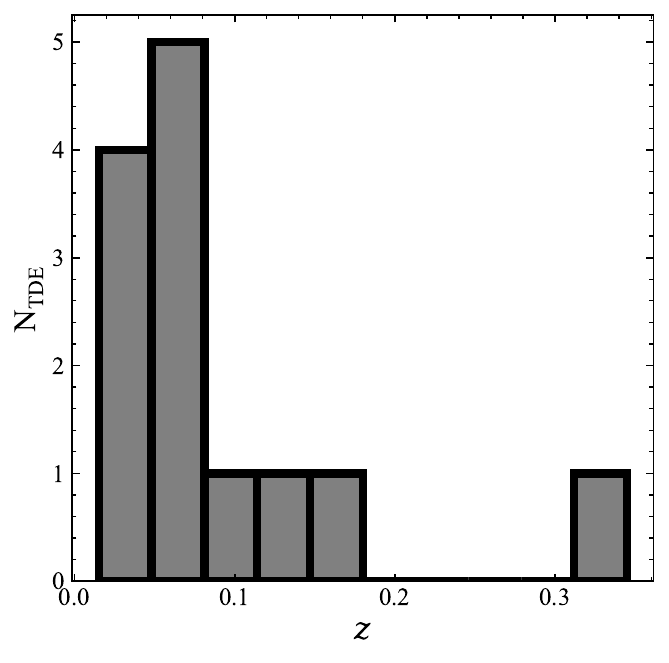}
    \caption{The distribution of redshifts for the TDE host galaxies in our sample. The distribution peaks below $z \sim 0.1$, with the highest redshift object, AT2020qhs, at $z=0.345$. Values are taken from \citet{vanVelzen21} and \citet{Hammerstein23}.}
    \label{fig:redshift}
\end{figure}

We selected our host galaxy sample from the ZTF-I TDEs published in \citet{vanVelzen21} and \citet{Hammerstein23}, with the intention of constructing a sample which includes multiple TDE spectral classes and X-ray brightnesses. We point to \citet{vanVelzen21} for a full description of the ZTF TDE search, although we note that the method for discovering TDEs is agnostic to host galaxy type apart from filtering out known AGN. While this search is thus agnostic to host galaxy type, we do note that our selection of TDE hosts from the ZTF sample, designed to include TDEs from all classifications, will not follow the true observed rate of each type of TDE. However, this is likely not relevant for the study presented here as we do not make conclusions by comparing the TDE types. We show SDSS and Pan-STARRS images of each of the host galaxies in Figure \ref{fig:hosts}.
Our sample of thirteen TDEs includes all four TDE spectral classes \citep[for a description of all classes, see][]{Hammerstein23}, with 2 TDE-H, 8 TDE-H+He, 2 TDE-He, and 1 TDE-featureless, 6 of which are also X-ray detected TDEs. The hosts span redshifts in the range $0.015 \leq z \leq 0.345$ and have stellar masses in the range $9.56 \leq \log(M_{\rm gal}/M_\odot) \leq 11.23$, both of which we take from the published values of \citet{vanVelzen21} and \citet{Hammerstein23}. In Figure \ref{fig:redshift}, we show the redshift distribution of the TDE hosts. In Sections \ref{sec:bhmass} and \ref{sec:kinematic}, we separate and discuss our results based on resolution.

\begin{figure}
    \centering
    \includegraphics[width=\columnwidth]{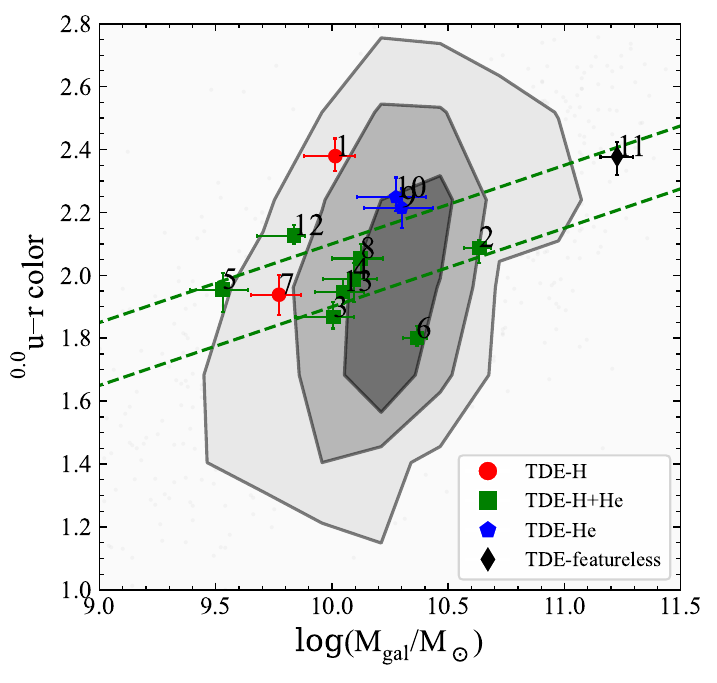}
    \caption{The rest-frame, extinction corrected $u-r$ color as a function of host galaxy mass for the TDE host galaxies and a sample of 955 galaxies from the SAMI survey. The dashed green lines indicate the location of the green valley, the location of which we take from \citet{Hammerstein23}. The colors and shapes of the points indicate the spectral class of TDE for each event. IDs are listed in Table \ref{tab:sample}. The TDE hosts are typically less massive than the background sample and more often reside in the green valley compared to the background galaxies ($\sim$54\% vs. $\sim$20\%).}
    \label{fig:ur}
\end{figure}

In Figure \ref{fig:ur} we show the rest-frame, extinction corrected $u-r$ color from \citet{Hammerstein23} derived from fitting the host SED for the TDE host galaxies as a function of host galaxy stellar mass. We also include a background sample of 955 galaxies from the SAMI Galaxy Survey DR3 \citep{Croom2021}, which provides spatially resolved stellar kinematic and population information, discussed further in Section \ref{sec:kinematic}. The galaxies in the SAMI sample were selected to span the plane of mass and environments, with the redshifts spanning $0.004 \leq z \leq 0.095$, masses between $10^7 - 10^12 M_\odot$, magnitudes with $r_{\rm pet} < 19.4$, and environments from isolated galaxies to groups and clusters \citep{Bryant2015}. $\sim$54\% of the TDE hosts are in the green valley compared to just $\sim$20\% of the background galaxies, in line with previous findings \citep[e.g.,][]{Hammerstein2021, Sazonov2021, Hammerstein23, Yao2023}. We summarize the properties of the host galaxies and include references to the first TDE classification in Table \ref{tab:sample}.

\begin{deluxetable*}{c l c c c c c D c c}
\tablecaption{Sample of TDE Host Galaxies}
\tablehead{\colhead{ID} & \colhead{Name} & \colhead{RA} & \colhead{Dec.} & \colhead{First TDE Classification} & \colhead{Spectral Class} & \colhead{Redshift} & \multicolumn2c{$\log(M_{\rm gal}/M_\odot)$} & \colhead{$m_r$} & \thead{$\sigma_{\rm instr}$ \\ (km $s^{-1}$)}}
\decimals
\startdata
1 & \textbf{AT2018zr} & 07:56:54.55 & +34:15:43.6 &\citet{AT2018zr} & TDE-H & 0.071 & $10.01_{-0.14}^{+0.08}$ & 18.02 & 18.3\\
2 & AT2018bsi & 08:15:26.63 & +45:35:32.0 & \citet{AT2018bsi} & TDE-H+He & 0.051 & $10.62_{-0.07}^{+0.05}$ & 15.50 & 18.8\\
3 & \textbf{AT2018hyz} & 10:06:50.88 & +01:41:33.9 & \citet{AT2018hyz} & TDE-H+He & 0.046 & $9.96_{-0.16}^{+0.09}$ & 16.96 & 16.3\\
4 & AT2018lni & 04:09:37.65 & +73:53:41.7 & \citet{vanVelzen21} & TDE-H+He & 0.138 & $10.10_{-0.13}^{+0.10}$ & 19.46 & 15.4\\
5 & AT2018lna & 07:03:18.65 & +23:01:44.7 & \citet{AT2018lna} & TDE-H+He & 0.091 & $9.56_{-0.14}^{+0.11}$ & 19.51 & 17.1\\
6 & \textbf{AT2019azh} & 08:13:16.95 & +22:38:53.9 & \citet{AT2019azh}\tablenotemark{a} & TDE-H+He & 0.022 & $9.74_{-0.05}^{+0.08}$ & 14.39 & 22.1 \\
7 & \textbf{AT2019ehz} & 14:09:41.91 & +55:29:27.8 & \citet{AT2019ehz} & TDE-H & 0.074 & $9.81_{-0.12}^{+0.09}$ & 18.72 & 19.8 \\
8 & AT2019qiz & 04:46:37.88 & $-$10:13:34.9 & \citet{AT2019qiz} & TDE-H+He & 0.015 & $10.01_{-0.12}^{+0.10}$ & 14.17 & 18.6 \\
9 & \textbf{AT2020ddv} & 09:58:33.42 & +46:54:40.4 & \citet{AT2020ddv} & TDE-He & 0.160 & $10.30_{-0.16}^{+0.13}$ & 19.37 & 14.9 \\
10 & \textbf{AT2020ocn} & 13:53:53.80 & +53:59:49.7 &\citet{AT2020ocn} & TDE-He & 0.070 & $10.28_{-0.17}^{+0.13}$ & 17.57 & 18.3 \\
11 & AT2020qhs & 02:17:53.95 & $-$09:36:50.9 & \citet{Hammerstein23} & TDE-featureless & 0.345 & $11.23_{-0.07}^{+0.07}$ & 19.40 & 13.0 \\
12 & AT2020wey & 09:05:25.91 & +61:48:09.1 & \citet{AT2020wey} & TDE-H+He & 0.027 & $9.63_{-0.22}^{+0.18}$ & 16.61 & 22.1 \\
13 & AT2020zso & 22:22:17.13 & $-$07:15:58.9 & \citet{AT2020zso} & TDE-H+He & 0.057 & $10.05_{-0.12}^{+0.09}$ & 17.03 & 21.4 \\
\enddata
\label{tab:sample}
\tablecomments{Labels used in figures, RA and Dec, TDE classification references, spectral classes, redshifts, host galaxy stellar masses, and host galaxy apparent $r$-band magnitudes for the thirteen objects in our sample. All spectral classifications, redshifts, and host galaxy stellar masses are based on those provided in \citet{vanVelzen21} and \citet{Hammerstein23}. Host magnitudes are derived from Pan-STARRS. X-ray detected events are bolded. We also provide the instrumental resolution, $\sigma_{\rm instr}$, measured from the FWHM of the arc spectrum at the observed wavelength of the \ion{Ca}{2} H and K lines for each object.}
\tablenotetext{a}{See also \citet{Hinkle2021}.}
\end{deluxetable*}

\subsection{Large Monolithic Imager and GALFIT} \label{sec:lmi}
We obtained optical imaging of the host galaxies in our sample using the Large Monolithic Imager (LMI) mounted on the 4.3-m Lowell Discovery Telescope (LDT) in Happy Jack, AZ. Data were obtained on 2022-10-30, 2022-11-30, and 2023-02-13 (PIs: Hammerstein, O'Connor) under clear skies and good observing conditions (seeing $\sim$1\arcsec). The targets were observed in the SDSS $r$-band filter with varying exposure times depending on the galaxy brightness, e.g., from 50 s for $r\approx14$ AB mag to 200 s for $r\approx19.5$ AB mag. The chosen exposure times lead to a high signal-to-noise ratio (SNR) for each galaxy, which when combined with the spatial resolution of LMI allow for an improved morphological analysis when compared to available archival data (e.g., SDSS). We were able to observe all thirteen host galaxies through this program. We reduced the LMI data using a custom \texttt{python} pipeline \citep[see][]{Toy2016,OConnor2022} to perform bias subtraction, flat-fielding, and cosmic ray rejection. The observations for each galaxy, including observation date, exposure time, and seeing during each observation are described in Table \ref{tab:lmi}. Given that the LMI observations were obtained several years after peak for all objects, we do not expect that the transient will contribute any appreciable flux to the photometry that may affect the fitting performed here.

\begin{figure*}
    \centering
    \includegraphics[width=\textwidth]{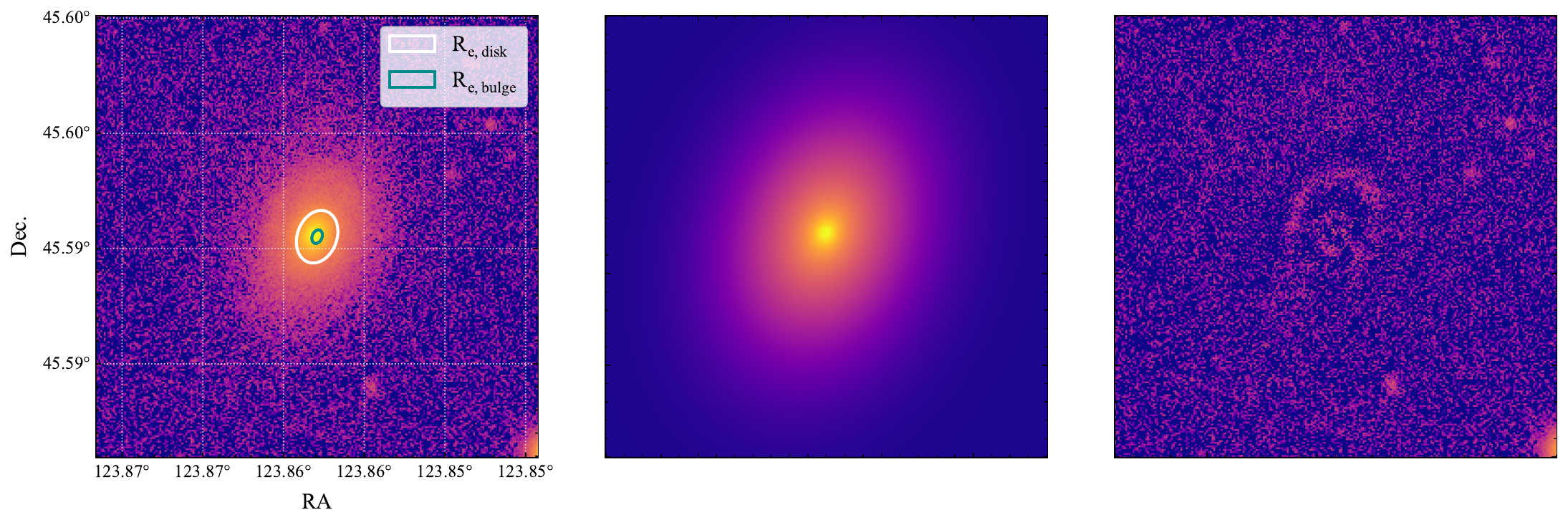}
\caption{A 29\arcsec$\times$29\arcsec~cutout of the LMI observations of the host galaxy of AT2018bsi, shown with the \galfit~model and residuals. All images are on the same scale. \galfit~is able to model the host galaxy reasonably well with the residuals showing potential dust lane or spiral arm features which are not as straightforward to model with \galfit~and for the purposes of the study presented here, are unimportant. In the left panel we show two  ellipses representing the fitted bulge effective radius ($R_{\rm e,bulge}$, cyan) and the disk effective radius (where the relationship between the effective radius and the scale length of the disk is $R_{\rm e,disk} = 1.678 R_{\rm s,disk}$, white).}
    \label{fig:galfit}
\end{figure*}

We use \texttt{GALFIT} \citep{Peng2002} to perform 2D fits to the host galaxy photometry and obtain morphological parameters such as the effective radius, ellipticity, and position angle of the host galaxies. Because we are interested in exploring galaxy properties at several different scales, we perform two fits with two different models. The first model includes a S\'ersic component and exponential disk component which is used to obtain a bulge effective radius ($R_{\rm e,bulge}$). This radius is used to mask a region in the IFU data for obtaining the bulge velocity dispersion and subsequently the black hole mass. The second fit includes a single S\'ersic component, used to obtain the effective radius of the entire galaxy light profile ($R_{\rm e,gal}$). We use this radius to mask the region for general kinematic and stellar population analysis. We fit all galaxies using these two models with the exception of AT2019qiz. The prominent bar in AT2019qiz required the addition of another component in order to isolate the bulge of the galaxy. Instead, we used a model which includes an exponential disk and two S\'ersic components, one for the bulge and one for the bar, which was sufficient to isolate the bulge and obtain the bulge effective radius. Some galaxies required additional components to mask out nearby stars or faint galaxies in the fitting window, which we included when necessary. We present the results of this fitting, namely the galaxy and bulge effective radii, in Table \ref{tab:results} and show an example fit and residuals in Figure \ref{fig:galfit}.

\begin{deluxetable}{l c r r}
\tablecaption{Summary of LMI observations}
\tablehead{ \colhead{Name} & \colhead{Obs. Date} & \thead{Exp. Time \\ (s)} & \thead{Seeing \\ \arcsec}}

\startdata
AT2018zr & 2022 Oct. 31 & 150 & 1.0 \\
AT2018bsi & 2022 Dec. 01 & 55 & 1.0 \\
AT2018hyz & 2022 Dec. 01 & 80 & 1.1 \\
AT2018lni & 2022 Dec. 01 & 200 & 1.1 \\
AT2018lna & 2022 Oct. 31 & 200 & 1.1 \\
AT2019azh & 2022 Oct. 31 & 70 & 1.0 \\
AT2019ehz & 2023 Feb. 13 & 120 & 1.9 \\
AT2019qiz & 2022 Dec. 01 & 50 & 1.1 \\
AT2020ddv & 2022 Oct. 31 & 200 & 1.3 \\
AT2020ocn & 2022 Dec. 01 & 100 & 1.1 \\
AT2020qhs & 2022 Dec. 01 & 200 & 1.0 \\
AT2020wey & 2022 Oct. 31 & 80 & 1.4 \\
AT2020zso & 2022 Dec. 01 & 60 & 1.2\\
\enddata
\label{tab:lmi}
\tablecomments{Summary of observations obtained with LMI, including the observation date, exposure time, and seeing measured from the PSF of the observation. All observations were performed using the SDSS $r$-band filter.}
\end{deluxetable}

\subsection{Keck Cosmic Web Imager and GIST} \label{sec:kcwi}
We present Keck Cosmic Web Imager \citep[KCWI;][]{Morrissey2018} observations of thirteen TDE host galaxies selected from the ZTF-I sample of TDEs. Integral field spectra were obtained on the night of 2021-12-25 under clear weather conditions (seeing $\sim 0.8\arcsec$) as part of program ID N096 (PI: Gezari). Observations for each object, described in Table \ref{tab:kcwi}, were obtained using the small ($8\farcs4 \times 20\farcs4$) slicer and `BM' grating, which gives a nominal resolution of $R_{\rm 0}=8000$ and an average bandpass of 861 \AA. In Table \ref{tab:kcwi}, we provide the instrumental resolution, $\sigma_{\rm instr}$, for each object measured from the FWHM of the arc spectrum at the observed wavelength of the \ion{Ca}{2} H and K lines. We also provide the days since peak for each observation as well as the average seeing between coadded exposures in Table \ref{tab:kcwi}. Three different central wavelengths were used to ensure that important host galaxy stellar absorption lines were observed for each galaxy. The final configurations are as follows:
\begin{itemize}
    \item[\it i.] {\it C1}: Small slicer, `BM' grating, central wavelength of 4200 \AA. 
    \item[\it ii.] {\it C2}: Small slicer, `BM' grating, central wavelength of 4800 \AA.
    \item[\it iii.] {\it C3}: Small slicer, `BM' grating, central wavelength of 5200 \AA.
\end{itemize}

In Figure \ref{fig:hosts}, we overplot the KCWI pointing for each observed galaxy. Three host galaxies, AT2018bsi, AT2019azh, and AT2019qiz, have angular sizes larger than the KCWI field-of-view. For each of these galaxies we obtained sky exposures offset from the host galaxy in order to perform sky subtraction.

The observations were reduced using the standard procedure of the KCWI data reduction pipeline \citep{Neill2023} which includes bias subtraction, flat fielding, cosmic ray removal, sky subtraction, wavelength calibration, heliocentric correction, and flux calibration. We used \texttt{CWITools} \citep{OSullivan20} to apply a WCS correction to the KCWI data in `src\_fit' mode, which fits 1D profiles to the spatial data to find the peak of the source and then applies a correction to the WCS such that the peak aligns with the input coordinates.

We use the Galaxy IFU Spectroscopy Tool \citep[\gist;][]{Bittner2019} modified to work with KCWI data to obtain the stellar kinematic and population information. The \gist~pipeline performs all necessary steps to analyze the KCWI IFU spectra with \ppxf~\citep{Cappellari2022}, including spatial masking and binning, SNR determination and masking, stellar kinematic analysis, and stellar population analysis. The X-shooter library of simple stellar population models \citep[XSL;][]{Verro2022} offers the best spectral resolution ($\sigma \sim 13$ km s$^{-1}$, $R\sim10000$) and wavelength coverage (3500 \AA -- 24800 \AA) which matches our KCWI observations ($\lambda_{\rm obs,min}=3768$ \AA~in configuration C1 and $\lambda_{\rm obs,max}=5624$ \AA~in configuration C3), meaning we can fit the entire spectral range for each host galaxy. The XSL provides several options for initial mass functions (IMF) and isochrones. We choose the set of models that utilizes the Salpeter IMF \citep{salpeter55} and PARSEC/COLIBRI isochrones \citep{Bressan2012, Marigo2013}, which includes stellar populations with ages above 50 Myr and metallicities in the range $-2.2 <$ [Fe/H] $< +0.2$, normalized to obtain mass-weighted stellar population results.

We run the \gist~pipeline three times for each host galaxy, each time using different binning and masking criteria, and using 1000 Monte-Carlo simulations to extract the uncertainties on the stellar kinematics. We spatially mask and bin the spaxels for the three different fits as follows:
\begin{itemize}
    \item[\it i.] {\it Bulge $\sigma$ fit}: Mask all spaxels outside of $R_{\rm e,bulge}$ obtained from \galfit; combine remaining spaxels into one bin to obtain $\sigma$, the bulge velocity dispersion.
    \item[\it ii.] {\it Galaxy $(V/\sigma)_e$ fits}: Mask all spaxels outside of $R_{\rm e,gal}$ obtained from \galfit; apply no binning to obtain the spatially resolved galaxy line-of-sight velocities ($V$) and velocity dispersions ($\sigma$), with $(V/\sigma)_e$ being the ratio of random to ordered motion within the galaxy effective radius.
    \item[\it iii.] {\it Stellar population fit}: Mask all spaxels outside of $R_{\rm e,gal}$ obtained from \galfit; combine remaining spaxels into one bin.
\end{itemize}
We are motivated to perform three different fits for several reasons. The first is so that our black hole masses are determined only from the bulge velocity dispersions, with the bulge effective radius determined from the two component \texttt{GALFIT} fit. The second is so that our determination of the large-scale kinematics and stellar population properties follows most closely the methods of \citet{vandeSande2018}, who perform two fits within an ellipse that encloses half of the projected total galaxy light: one which is similar to our galaxy $(V/\sigma)_e$ fit and one which is similar to our stellar population fit. There are four cases in which the bulge effective radius is smaller than the seeing of the KCWI observations: AT2018lni, AT2020ddv, AT2020ocn, and AT2020qhs. For these objects, instead of simply using the bulge effective radius given by \galfit~to perform the bulge $\sigma$ fit, we use the sum in quadrature of the bulge effective radius and the seeing given in Table \ref{tab:kcwi}. The galaxy effective radius for AT2018lni is also smaller than the seeing, and in this case, we use the sum in quadrature of the galaxy effective radius and the seeing to perform the galaxy $(V/\sigma)_e$ fits and the stellar population fit. We present and discuss the results of this analysis in the next sections.

\begin{deluxetable}{l c c D c}
\tablecaption{Summary of KCWI observations}
\tablehead{ \colhead{Name} & \colhead{Config.} & \thead{Exp. Time \\ (s)} & \multicolumn2c{\thead{$\Delta t_{\rm obs - peak}$ \\ (days)}} & \thead{Seeing \\ \arcsec}}
\decimals
\startdata
AT2018zr & C1 & $2\times900$ & 1372 & 0.72 \\
AT2018bsi & C1 & $2\times150$ & 1362 & 0.65 \\
AT2018hyz & C1 & $2\times600$ & 1150 & 0.61 \\
AT2018lni & C2 & $2\times1800$ & 1097 & 0.69 \\
AT2018lna & C1 & $2\times1500$ & 1067 & 0.82 \\
AT2019azh & C1 & $2\times100$ & 1008 & 0.68 \\
AT2019ehz & C1 & $2\times1000$ & 960 & 0.65 \\
AT2019qiz & C1 & $2\times500$ & 807 & 0.95 \\
AT2020ddv & C2 & $2\times1500$ & 655 & 0.71 \\
AT2020ocn & C1 & $2\times600$ & 585 & 0.52 \\
AT2020qhs & C3 & 1350, 500 & 511 & 0.75 \\
AT2020wey & C1 & $2\times200$ & 418 & 0.74 \\
AT2020zso & C1 & 300, 600 & 386 & 0.66 \\
\enddata
\label{tab:kcwi}
\tablecomments{Summary of observations obtained with KCWI, including the instrument configuration, exposure times, days post-peak from the tidal disruption flare, and the average seeing for the coadded observations. $t_{\rm peak}$ is taken from \citet{Hammerstein23}. The configuration notation is described in Section \ref{sec:kcwi}.}
\end{deluxetable}

\section{Results} \label{sec:results}
\begin{figure*}
    \centering
    \includegraphics[width=\textwidth]{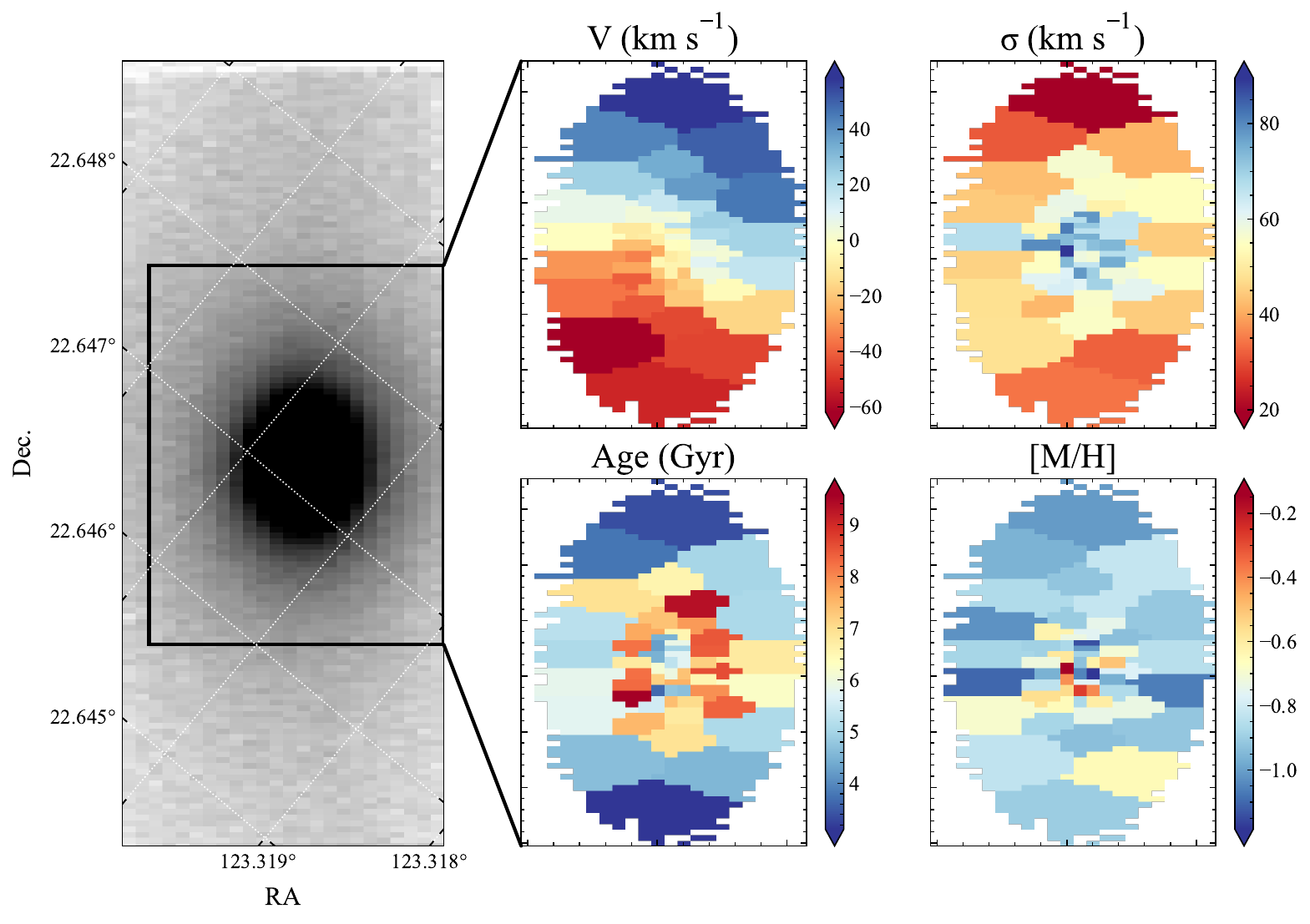}
    \caption{An example output from \gist~of the host galaxy of AT2019azh. The left panel shows an unbinned white light image of the KCWI observation. The panels on the right depict the output maps from \gist, which show the \ppxf-derived line-of-sight velocity, velocity dispersion, and stellar population ages and metallicities. The bins in this figure are constructed using the Voronoi binning method \citep{Cappellari2003} to reach a threshold SNR for each bin, in this case SNR $\sim 10$. We note that Voronoi binning is not performed for the fits used in the analysis.} This fit involves no spatial masking like that described in Section \ref{sec:kcwi}, but instead masks spaxels below the isophote level which has a mean SNR of 2.2. This particular fit is not used for any analysis and is for illustrative purposes only.
    \label{fig:gist}
\end{figure*}

\begin{figure*}
    \centering
    \includegraphics[width=\textwidth]{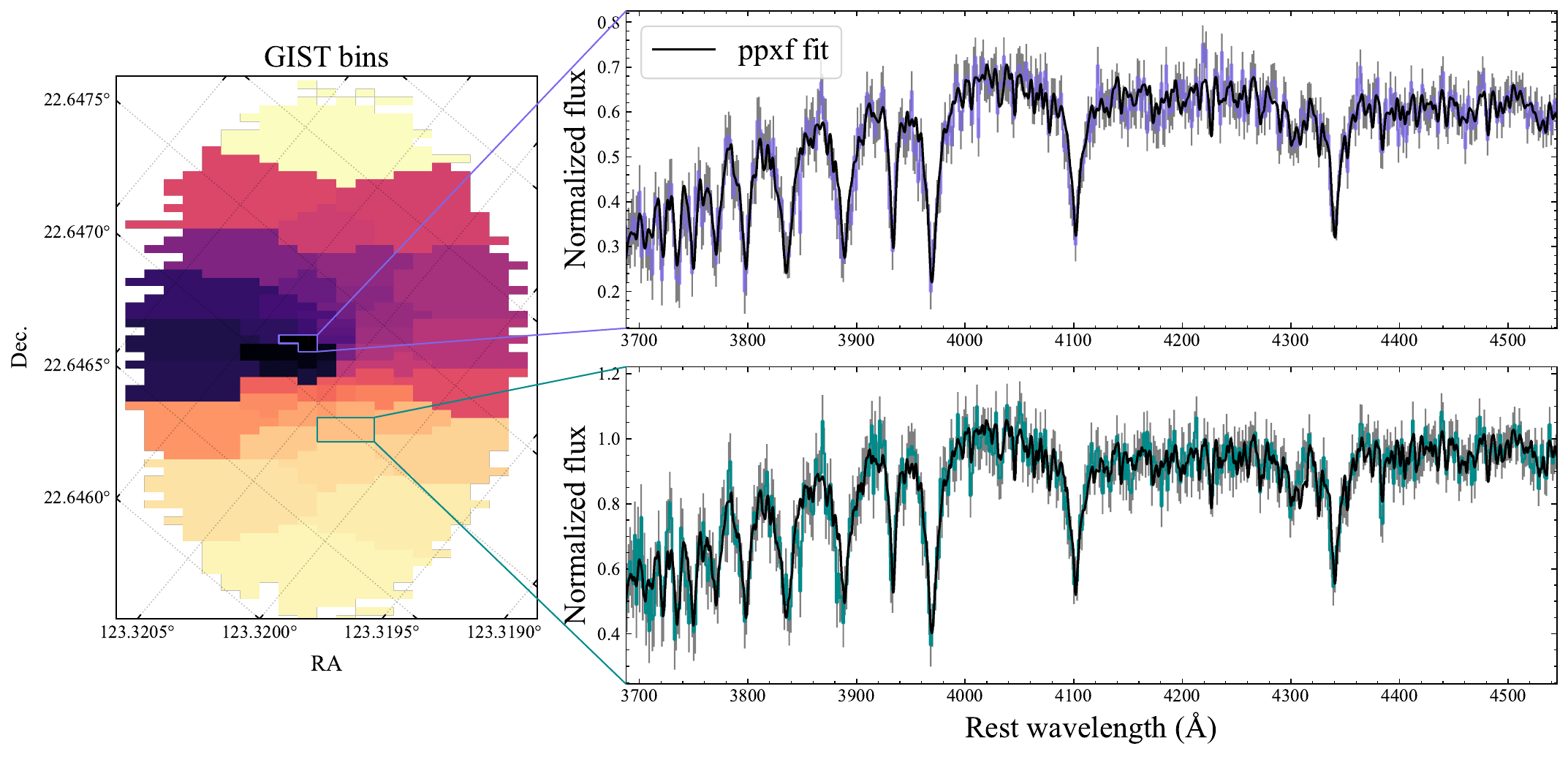}
    \caption{Example \ppxf~fits to the host galaxy of AT2019azh output from \gist. The left panel shows the bins constructed with \gist~where the color represents the bin to which each spaxel belongs. Bins are constructed using the Voronoi binning method \citep{Cappellari2003} to reach a threshold SNR for each bin, in this case SNR $\sim10$. We note that Voronoi binning is not performed for the fits used in the analysis. The two panels on the right show the spectra (purple and teal lines) and \ppxf~fits (black lines) from the outlined bins on the left. We show the uncertainties on the spectra in gray.}
    \label{fig:ppxf}
\end{figure*}

We present the results of our kinematic and stellar population analysis on the KCWI spectra of the 13 TDE host galaxies. We summarize our main results in Table \ref{tab:results}. In Figure \ref{fig:gist}, we show a white light image of the host galaxy of AT2019azh and example output maps from \gist, including the line-of-sight velocity and velocity dispersion as well as the stellar population age and metallicity. In Figure \ref{fig:ppxf}, we show the bins constructed by \gist, as well as two example spectra and \ppxf~fits from different bins. The output we show in Figures \ref{fig:gist} and \ref{fig:ppxf} involves no spatial masking like that described in Section \ref{sec:kcwi}, but instead masks spaxels below the isophote level which has a mean SNR of 2.2. This particular fit is not used for any analysis and is for illustrative purposes only.

One important comparison to make for all results is that of the differing angular resolutions resulting from the range of redshifts for the TDE hosts. As such, we investigate whether angular resolution may influence the results we discuss in Sections \ref{sec:bhmass} and \ref{sec:kinematic}. We split our sample into three different angular resolution bins:
\begin{itemize}
    \item[\it i.] {$\sim$0.5 kpc/\arcsec}: AT2019azh, AT2019qiz, AT2020wey
    \item[\it ii.] {$\sim$1.0 kpc/\arcsec}: AT2018bsi, AT2018hyz, AT2020zso
    \item[\it iii.] {$\gtrsim$1.3 kpc/\arcsec}: AT2018zr, AT2018lni, AT2018lna, AT2018ehz, AT2020ddv, AT2020ocn, AT2020qhs
\end{itemize}
We perform an Anderson-Darling test to compare these three subsamples and find that we cannot reject the null hypothesis that they are drawn from the same distribution of host galaxy stellar mass, velocity dispersion, black hole mass, or (V/$\sigma)_e$ ($p$-value $\geq 0.25$ for all tests). However, the sample sizes compared are small and may not provide a true measure of how angular resolution affects studies such as the one presented here. In the following sections, we discuss our results on obtaining the black hole masses and characterizing the host galaxy stellar kinematics and populations.

\begin{deluxetable*}{l r r r D D D r}
\tablecaption{Results from photometric and kinematic analysis}
\tablehead{ \colhead{Name} & \colhead{kpc$/\arcsec$} & \thead{$R_{\rm e, gal}$ \\ (\arcsec)} & \thead{$R_{\rm e,bulge}$ \\ (\arcsec)} & \multicolumn2c{$\sigma_\star$ (km s$^{-1}$)} & \multicolumn2c{$\log(M_{\rm BH} / M_\odot)$} & \multicolumn2c{(V/$\sigma)_e$} & \thead{Age \\ (Gyr)}}
\decimals
\startdata
AT2018zr  & 1.35 & 1.87 & 0.89 & 49.79 $\pm$4.93 & 5.56 $\pm$ 0.76 & 0.52 $\pm$ 0.20 & 2.65\\
AT2018bsi & 1.00 & 6.15 & 1.84 & 117.54 $\pm$8.12 & 7.14 $\pm$ 0.62 & 0.93 $\pm$ 0.15 & 0.57\\
AT2018hyz & 0.90 & 1.34 & 0.69 & 66.62 $\pm$3.12 & 6.10  $ \pm$  0.67 & 0.12  $ \pm$  0.05 & 6.95\\
AT2018lni & 2.44 & 0.56 (0.88) & 0.34 (0.78) & 59.47 $\pm$3.78 & 5.89  $ \pm$  0.70 & 0.26  $ \pm$  0.09 & 8.65\\
AT2018lna & 1.70 & 1.15 & 0.92 & 36.43 $\pm$4.52 & 4.98  $ \pm$  0.83 & 0.78  $ \pm$  0.38 & 3.23\\
AT2019azh & 0.45 & 9.75 & 2.52 & 68.01 $\pm$2.02 & 6.13  $ \pm$  0.66 & 0.88  $ \pm$  0.11 & 8.68\\
AT2019ehz & 1.41 & 1.76 & 1.15 & 46.65 $\pm$11.83 & 5.44  $ \pm$  0.98 & 0.37  $ \pm$  0.20 & 6.03\\
AT2019qiz & 0.31 & 8.85 & 2.27 & 71.85 $\pm$1.93 & 6.23  $ \pm$  0.65 & 0.71  $ \pm$  0.08 & 2.15\\
AT2020ddv & 2.76 & 0.88 & 0.47 (0.85) & 73.44 $\pm$10.06 & 6.28  $ \pm$  0.78 & 0.09  $ \pm$  1.11 & 6.12\\
AT2020ocn & 1.34 & 1.40 & 0.28 (0.59) & 90.15 $\pm$4.46 & 6.65  $ \pm$  0.63 & 0.36  $ \pm$  0.14 & 8.09\\
AT2020qhs & 4.89 & 2.05 & 0.72 (1.04) & 188.69 $\pm$37.86 & 8.01  $ \pm$  0.82 & 0.53  $ \pm$  0.15 & 1.98\\
AT2020wey & 0.55 & 2.49 & 0.87 & 53.54 $\pm$4.75 & 5.69  $ \pm$  0.74 & 0.40  $ \pm$  0.32 & 8.43\\
AT2020zso & 1.10 & 2.57 & 1.08 & 61.80 $\pm$4.93 & 5.96  $ \pm$  0.71 & 1.08  $ \pm$  0.27 & 6.32
\enddata
\label{tab:results}
\tablecomments{The results from our photometric and kinematic analysis of the LMI and KCWI data, including the galaxy and bulge half light radii measured from \galfit, the bulge velocity dispersion and derived black hole mass, the ratio of ordered rotation to random stellar motion (V/$\sigma)_e$, and the stellar population age within the galaxy effective radius. For AT2018lni, AT2020ddv, AT2020ocn, and AT2020qhs, the values in parentheses are the values obtained from adding the \galfit~values and the KCWI seeing in quadrature, and are the values used to extract the bulge $\sigma$ fits, and in the case of AT2018lni, the galaxy kinematics and stellar population fits.}
\end{deluxetable*}

\section{Black hole masses} \label{sec:bhmass}
We derive the black hole masses through the \msigma~relation of \citet{gultekin09}, assuming that this relation holds valid for all galaxies in this sample:
\begin{equation}
\log(M_{\rm BH}/M_\odot) = 8.12 + 4.24\log \left(\frac{\sigma}{200~\mathrm{km~s^{-1}}} \right )
\end{equation}
We propagate the uncertainties on the velocity dispersion through this relation and add them linearly with the intrinsic scatter on the relation to obtain the uncertainty on the black hole mass.

\begin{figure}
    \centering
    \includegraphics[width=\columnwidth]{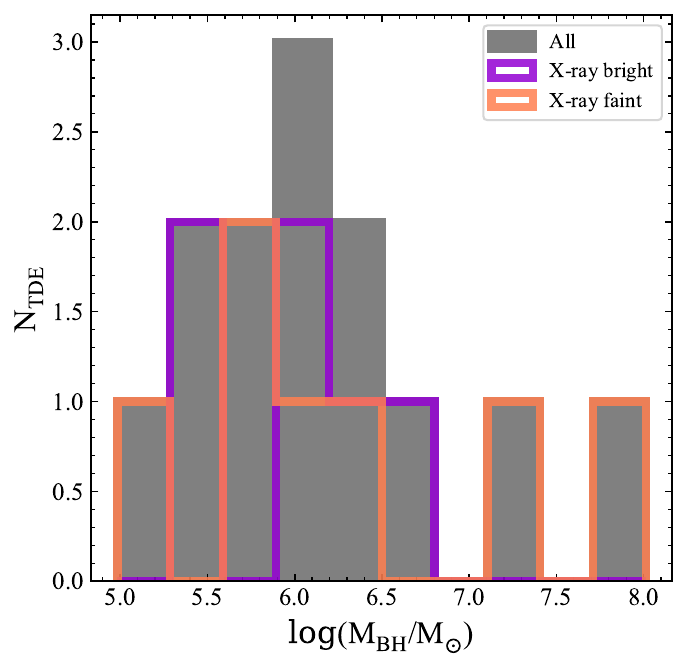}
    \caption{Distribution of black hole masses for the host galaxies in our sample. We show the entire sample in black, with the divisions on X-ray bright vs. X-ray faint in purple and orange, respectively. The distribution peaks at $\log(M_{\rm BH}/M_\odot) = 6.05$, consistent with previous results for similar analyses. We find no significant difference in black hole masses between the X-ray bright (6 total) and X-ray faint (7 total) events.}
    \label{fig:mbh_hist}
\end{figure}

In Figure \ref{fig:mbh_hist}, we show the distribution of black hole masses for the entire sample in addition to the subsamples of X-ray bright and X-ray faint events. We find that the distribution peaks at $\log(M_{\rm BH}/M_\odot) = 6.05$ with a range of masses $4.98 \leq \log(M_{\rm BH}/M_\odot) \leq 8.01$, which is consistent with previous studies performing a similar analysis \citep[e.g.,][]{Wevers2017, Wevers2019, Yao2023}. We examine whether the populations of X-ray bright and X-ray faint events show any significant difference in their black hole mass distributions by performing an Anderson-Darling test and find that we cannot reject the null hypothesis that the X-ray bright and X-ray faint samples are drawn from the same distribution in black hole mass ($p$-value $\geq 0.25$). This is consistent with several previous studies \citep[e.g.,][]{Wevers2019, French2020, Hammerstein23} which largely found no significant difference in the black hole, host galaxy, or even light curve properties between X-ray bright and X-ray faint TDEs. This lack of difference between X-ray bright and X-ray faint populations may be explained by the unifying theory of \citet{Dai2018}, which posits that whether or not X-rays are observed in a particular TDE is a matter of viewing angle effects.

\begin{figure}
    \centering
    \includegraphics[width=\columnwidth]{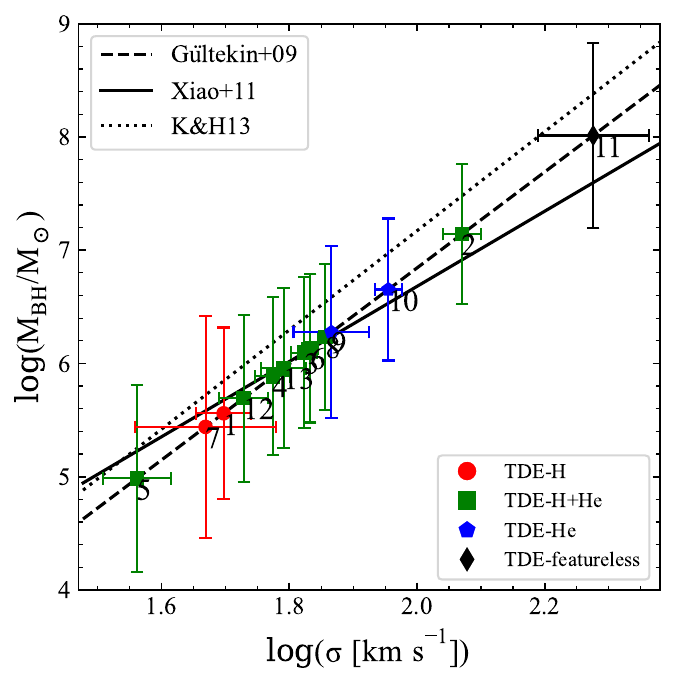}
    \caption{The black hole mass as a function of the velocity dispersion, along with several derived relations from the literature. We employ the relation of \citet{gultekin09} (G\"ultekin+09) to derive the black hole masses presented here. Black hole masses derived from \citet{kormendy13} (K\&H13) would generally be higher than those derived from \citet{gultekin09}, while the \citet{xiao11} relation (Xiao+11) would yield lower masses at the higher velocity dispersion end of the relation and higher masses at the lower velocity dispersion end of the relation. Labels for each TDE are in Table \ref{tab:sample}.}
    \label{fig:msigma}
\end{figure}

Figure \ref{fig:msigma} shows the black hole mass as a function of the velocity dispersion along with several derived relations from the literature, including \citet{gultekin09}, \citet{xiao11}, and \citet{kormendy13}. While values derived from the \citet{kormendy13} relation would generally be higher than those derived from the \citet{gultekin09} relation, the \citet{xiao11} relation is flatter, with higher velocity dispersion values yielding lower black hole masses and lower velocity dispersion values yielding higher black hole masses. We discuss further implications of our choice of \msigma~relation used to derive black hole masses in Sections \ref{sec:corr} and \ref{sec:AT2020qhs}.

\begin{figure}
    \centering
    \includegraphics[width=\columnwidth]{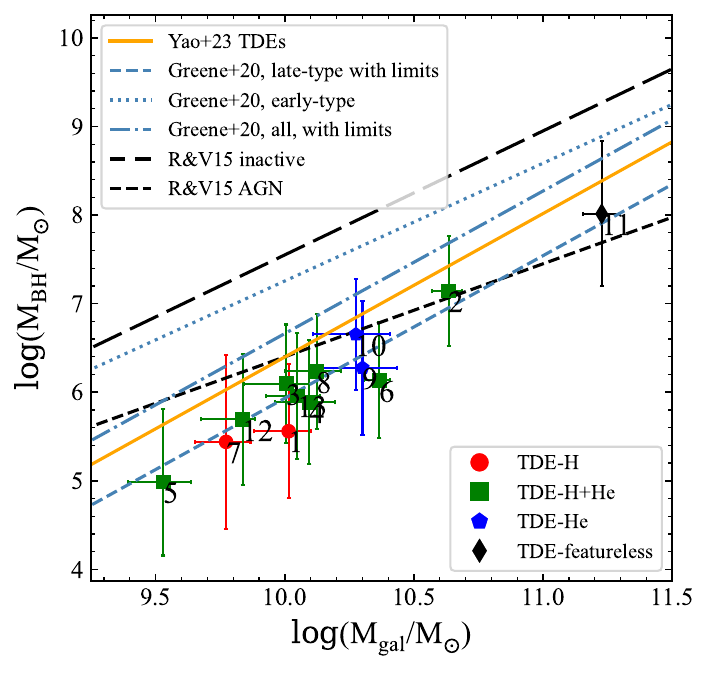}
    \caption{The black hole mass as a function of the host galaxy stellar mass. We show several derived $M_{\rm BH} - M_{\rm gal}$ relations. Black dashed and long-dashed lines show the relations from \citet{reines15} derived from AGN host galaxies and inactive galaxies, respectively. The blue dashed, dotted, and dot-dashed lines show the relations from \citet{Greene20} derived from late-type galaxies, early-type galaxies, and all galaxy types, respectively. We also showed the fitted relation from \citet{Yao2023}, which was fit only for TDE hosts. Labels for each TDE are in Table \ref{tab:sample}.}
    \label{fig:mbh_mgal}
\end{figure}

In Figure \ref{fig:mbh_mgal}, we show the derived black hole masses as a function of host galaxy stellar mass along with several empirical relations from the literature. \citet{reines15} derived the relations for AGN and inactive galaxies, while \citet{Greene20} derived the relations for late, early, and all galaxy types. Importantly, \citet{Greene20} used upper limits in their calculations which are crucial for including low-mass systems, such as the ones that host TDEs, in the relation. We also show the fitted relation from \citet{Yao2023}, which was derived by fitting a linear relation between $M_{\rm gal}$ and $M_{\rm BH}$ for the TDE hosts in their sample. Rather interestingly, the TDE hosts most closely follow the relation for late-type galaxies, despite very few being classified as such. This could be explained by the very few low-mass early-type galaxies used in deriving the relations for early-type galaxies and all galaxy types. Alternatively, this may be caused by our choice in \msigma~relation, although each scaling will have its own resulting offset.

\subsection{Comparisons to previous measurements}
All objects in our sample have previously measured black hole masses through a variety of methods, although only three have previously measured velocity dispersions. We compare our estimate of the black hole mass derived from the bulge velocity dispersion and \msigma~relation with previous estimates using the same method.

\textit{AT2019azh}: \citet{Yao2023} derived the black hole mass for AT2019azh by fitting the optical ESI spectrum using \ppxf. They found $\sigma_\star = 67.99 \pm 2.03$ km s$^{-1}$, corresponding to a black hole mass of $\log(M_{\rm BH}/M_\odot) = 6.44 \pm 0.33$ using the \msigma~relation of \citet{kormendy13}. Our value of $\sigma_\star = 68.01 \pm 2.02$ km s$^{-1}$ is consistent with that of \citet{Yao2023}.

\textit{AT2020wey}: \citet{Yao2023} also measured the velocity dispersion of the host galaxy of AT2020wey in the same manner as AT2019azh, finding $\sigma_\star = 39.36 \pm 2.79$ km s$^{-1}$. We find a significantly higher value for the velocity dispersion of $\sigma_\star = 53.54 \pm 4.75$ km s$^{-1}$. It is possible that with the small effective radius of AT2020wey (see Table \ref{tab:results}), the long-slit spectra used to derive the velocity dispersion in \citet{Yao2023} are inclusive of stars much farther from the bulge effective radius and thus have lower velocity dispersions. This may explain the discrepancy we see here. Indeed, a fit to the entire host galaxy of AT2020wey reveals that regions away from the nucleus have much lower velocity dispersions ($\sim 24$ km s$^{-1}$) which may influence the resulting black hole mass derived from stellar kinematics.

\textit{AT2019qiz}: \citet{nicholl20} fit the late time X-shooter spectrum of AT2019qiz using \ppxf~and found $\sigma_\star = 69.7 \pm 2.3$ km s$^{-1}$. Our value for the velocity dispersion is marginally higher, $\sigma_\star = 71.85 \pm 1.93$ km s$^{-1}$, but still consistent within the mutual uncertainties of the two measurements.

All objects in our sample also have at least one estimate of the black hole mass obtained from fitting the TDE light curve with the \texttt{MOSFiT} \citep{Guillochon2018} TDE model \citep{Mockler2019}. The TDE model fits each TDE by generating bolometric light curves via hydrodynamical simulations and passing them through viscosity and reprocessing transformation functions to create the the single-band, observed light curves. \texttt{MOSFiT} then uses the single-band light curves to fit the multi-band input data to estimate the light-curve properties and information on the disrupted star in addition to the mass of the SMBH. \citet{Hammerstein23} used \texttt{MOSFiT} to fit the light curves of every object in our sample, but found no significant correlation with the host galaxy mass.

\begin{figure}
    \centering
    \includegraphics[width=\columnwidth]{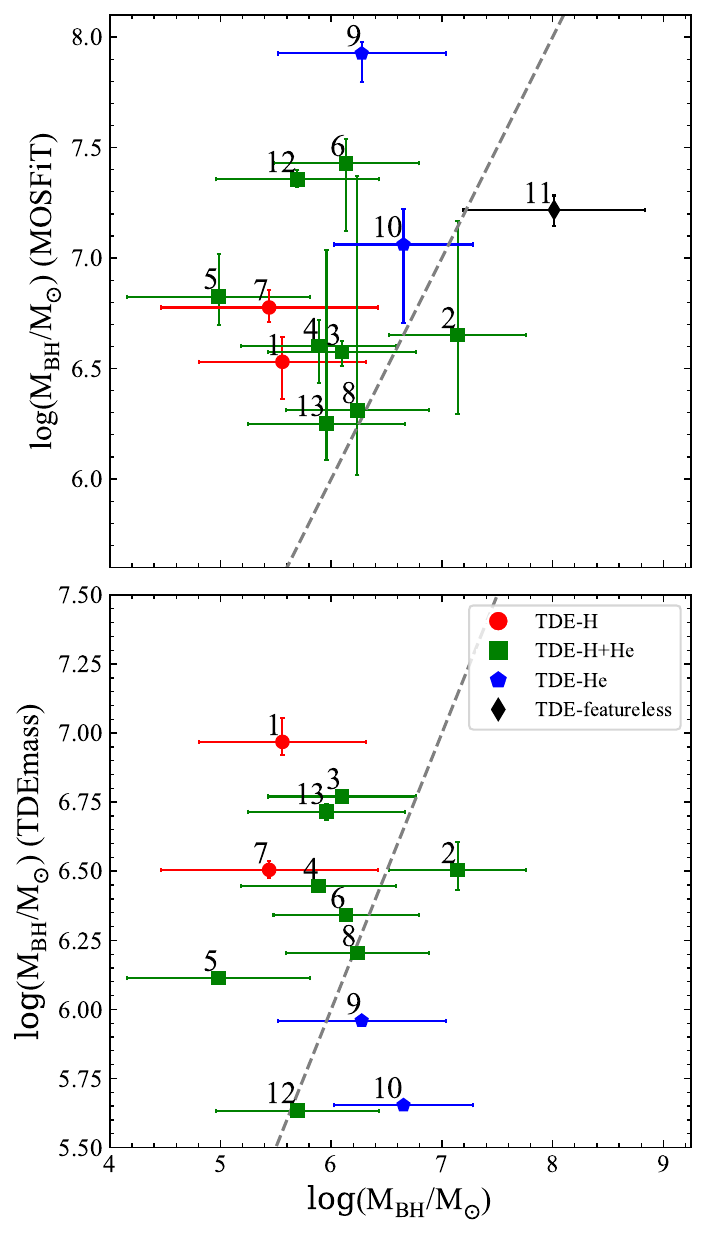}
    \caption{\textit{Top panel}: The black hole mass derived from \texttt{MOSFiT} as a function of the black hole mass we derive from host kinematics. The gray dashed line indicates a one-to-one relationship. We do not find a significant correlation between the two measurements. \textit{Bottom panel}: The black hole mass derived from \texttt{TDEmass} as a function of the black hole mass we derive from host kinematics. The gray dashed line indicates a one-to-one relationship. We note that the mass for AT2020qhs (ID 11) was not able to be determined with \texttt{TDEmass}. We do not find a significant correlation between the two measurements. Labels for each TDE are in Table \ref{tab:sample}.}
    \label{fig:mosfittdemass}
\end{figure}

We now reexamine any potential correlation using the derived black hole mass instead. In Figure \ref{fig:mosfittdemass}, we show the \texttt{MOSFiT} black hole mass as a function of the black hole mass we have derived here. The gray dashed line indicates a one-to-one relationship. While we do find a weak positive correlation between the \texttt{MOSFiT} masses and the masses we derive here using a Kendall's tau test ($\tau = 0.05$), it is not significant ($p$-value $=0.9$). As our \msigma~derived black hole masses are so well correlated with the host galaxy masses from \citet{Hammerstein23}, it is not surprising that we do not find a significant correlation between the \texttt{MOSFiT} masses and our masses. Given that the \texttt{MOSFiT} mass are typically orders of magnitude larger than those inferred through the \msigma~relation, it is possible that an underestimation of the black hole mass due to uncertainties of the relation at such low velocity dispersions is causing the discrepancy. Additional updates to the \texttt{MOSFiT} TDE model, which will be presented in Mockler \& Nicholl et al.~(2023, in prep), may also help to address the discrepancies.

\citet{Hammerstein23} also estimated the black hole mass using the \texttt{TDEmass} code \citep{Ryu2020} which assumes that circularization happens slowly, and that the UV/optical emission arises from shocks in the intersecting debris streams instead of in an outflow or wind. Again, they found no significant correlation between the SMBH mass and the host galaxy mass. We show the \texttt{TDEmass} SMBH mass as a function of the SMBH mass derived from stellar kinematics in Figure \ref{fig:mosfittdemass}, with gray dashed line indicates a one-to-one relationship. We note that the mass for AT2020qhs (ID 11) was not able to be determined with \texttt{TDEmass}. We find no significant correlation between the \texttt{TDEmass} values for the black hole mass and the ones we derive here ($p$-value $=0.4$).

While it is not surprising that the \texttt{MOSFiT} and \texttt{TDEmass} values do not agree, as they derive the black hole mass using differing assumptions on the origin of the UV/optical emission, the lack of any correlation with host galaxy properties is puzzling. Previous studies \citep[e.g.,][]{Ramsden2022, Mockler2019} which derive the black hole mass from \texttt{MOSFiT} have found weak correlations between the SMBH mass and properties such as the bulge mass and host galaxy stellar mass, but parameters such as the bulge mass can be difficult to determine for TDE host galaxies without sensitive imaging given their masses and redshifts. On the other hand, studies like \citet{Wevers2019} have confirmed a disparity between SMBH masses measured using \texttt{MOSFiT} and those from host scaling relations such as \msigma. The lack of correlation is not entirely discouraging, as there is indeed some correlation between light curve properties such as rise and fade timescale and the black hole mass \citep{vanVelzen21, Nicholl2022, Hammerstein23, Yao2023}, and perhaps indicates a need to revisit the exact ways in which the properties of the black hole are imprinted onto the observed TDE light curves.

\subsection{Correlations with TDE light curve properties} \label{sec:corr}
Many previous studies have found significant correlations between the light curve properties of TDEs and the black hole mass or, more often, the host galaxy mass. \citet{vanVelzen21} found a correlation between the decay timescale and host galaxy stellar mass, which \citet{Hammerstein23} further confirmed with a larger sample. This is consistent with many previous results in the literature \citep[e.g.,][]{Blagorodnova2017, Wevers2017}. \citet{Hammerstein23} additionally found a weak correlation between the rise timescale and the host galaxy stellar mass as well as between the peak luminosity and the host galaxy stellar mass.

We now reexamine the correlations with host galaxy stellar mass presented in \citet{Hammerstein23}. Between the SMBH mass and the decay rate for the 13 TDEs, we find a weak positive correlation with a Kendall's tau test, but the $\tau = 0.26$ correlation is not significant with $p$-value $= 0.25$. The Kendall's tau test between the SMBH mass and the rise results in $\tau = 0.41$, but again is not significant with a $p$-value $= 0.06$. We no longer find a correlation between the black hole mass and the peak blackbody luminosity. While we generally find the same trends as previous works, our smaller sample size weakens our ability to make significant conclusions and the disappearance of significant correlations here should be interpreted with caution.

\begin{figure}
    \centering
    \includegraphics[width=\columnwidth]{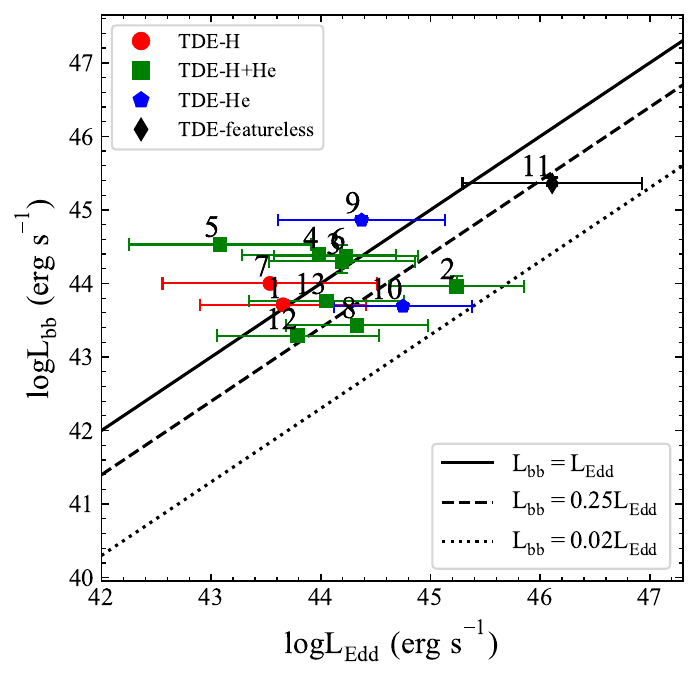}
    \caption{The peak blackbody luminosity as a function of the Eddington luminosity implied by the black hole mass. The solid, dashed, and dotted lines indicate constant Eddington ratios. We find that nearly all TDEs in our sample are consistent with being at or below the Eddington limit, with the exception of AT2018lna. This object has the lowest velocity dispersion in our sample and the black hole mass should be interpreted with caution. Labels for each TDE are in Table \ref{tab:sample}.}
    \label{fig:ledd}
\end{figure}

The black hole mass now makes it possible to compare the peak blackbody luminosity of the TDE light curves with the Eddington luminosity implied by the black hole mass. We define the Eddington luminosity as $L_{\rm Edd} \equiv 1.25 \times 10^{38} (M_{\rm BH}/M_\odot)$ and take values for the peak blackbody luminosity from \citet{Hammerstein23} measured using the peak UV/optical SED. In Figure \ref{fig:ledd}, we show the peak blackbody luminosity as a function of the Eddington luminosity, with solid, dashed, and dotted curves representing lines of constant Eddington ratio.

All of our events are consistent with being at or below the Eddington luminosity (solid line), apart from AT2018lna (ID 5), with its blackbody luminosity significantly super-Eddington even at the maximum extent of its uncertainties. We note that this is also the lowest mass object in our sample with $\log(M_{\rm BH}/M_\odot) = 4.98 \pm 0.83$. The apparent significantly super-Eddington luminosity may be due to the large uncertainty on the calibration of \msigma~relation at such low velocity dispersions, although without larger samples of dynamically measured masses for intermediate mass black holes, this problem is hard to constrain \citep[for a review on such measurements, see][]{Greene20}. If we instead obtain the mass for AT2018lna using the relation from \citet{xiao11}, derived from active galaxies with low black hole masses, we find that the resulting black hole mass is higher: $\log(M_{\rm BH}/M_\odot) = 5.22$. Although the peak luminosity is still super-Eddington. The mass for AT2018lna should thus be interpreted with caution. Super-Eddington mass fallback rates are not unexpected for black holes with such low masses, with duration of $\dot{M}/M_{\rm Edd} > 1$ longer for smaller black holes \citep{decolle12}. AT2018lna, the lowest mass black hole and the one with the largest Eddington ratio, does indeed follow this expected relation, its bolometric luminosity staying above Eddington for much longer than the other objects in this sample when examing the light curve fits of \citet{Hammerstein23}.

AT2020qhs is an outlier in black hole mass, but not necessarily an outlier in its Eddington ratio. \citet{Wevers2019} found that the TDE candidate ASASSN-15lh possessed similar qualities and that the observed emission is consistent with the peak Eddington ratio and luminosity of a maximally spinning Kerr black hole. As we discuss in Section \ref{sec:AT2020qhs}, a non-negligible spin may explain the properties of AT2020qhs.

\begin{figure}
    \centering
    \includegraphics[width=\columnwidth]{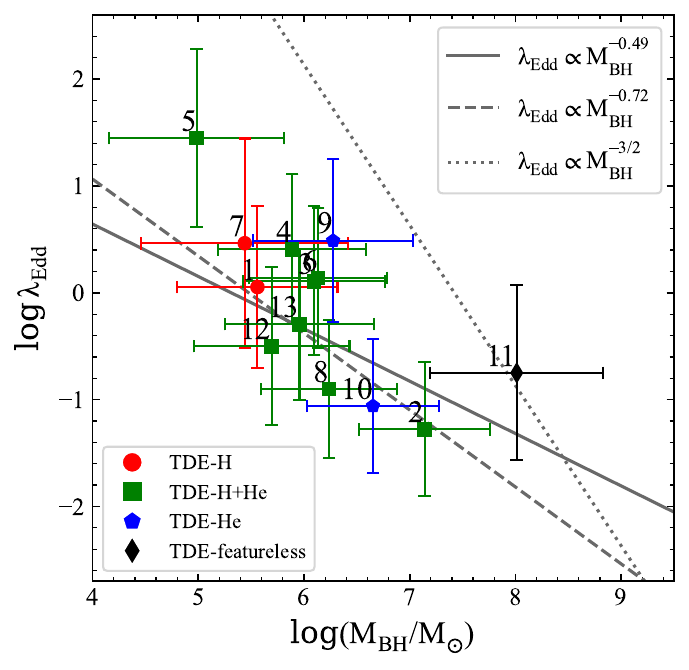}
    \caption{The Eddington ratio as a function of black hole mass. The dotted line is the expected Eddington ratio for the peak fallback accretion rate and the solid and dashed lines are the fitted relations from \citet{Yao2023} where $\dot{M}_{\rm fb}/\dot{M}_{\rm Edd} \propto M_{\rm BH}^{-0.49}$ and $\dot{M}_{\rm fb}/\dot{M}_{\rm Edd} \propto M_{\rm BH}^{-0.79}$, respectively. We find a moderate negative correlation between $\lambda_{\rm Edd}$ and the black hole mass, with the relation shallower than the expected $\lambda_{\rm Edd} \propto M_{\rm BH}^{-3/2}$, but likely steeper than that obtained by \citet{Yao2023}. Labels for each TDE are in Table \ref{tab:sample}.}
    \label{fig:fedd}
\end{figure}

\citet{Yao2023} found a correlation between the Eddington ratio ($\lambda_{\rm Edd} \equiv L_{\rm bb}/L_{\rm Edd}$) and the black hole mass which was inconsistent with the expected ratio between the peak fallback rate and Eddington accretion rate. Instead, they found a much shallower relation between $\dot{M}_{\rm fb}/\dot{M}_{\rm Edd}$ and the black hole mass, which they attribute to either Eddington-limited accretion or that the UV/optical luminosity only captures a fraction of the total bolometric luminosity. We report similar findings here, with a moderate negative correlation between $\lambda_{\rm Edd}$ and $M_{\rm BH}$ resulting from a Kendall's tau test ($\tau = -0.46$, $p$-value = 0.03). In Figure \ref{fig:fedd}, we show $\log(\lambda_{\rm Edd})$ as a function of $M_{\rm BH}$, along with the fitted relations from \citet{Yao2023} (solid line, fitted for all 33 TDEs in their sample: $\dot{M}_{\rm fb}/\dot{M}_{\rm Edd} \propto M_{\rm BH}^{-0.49}$, dashed line, correcting for selection bias by only fitting objects with $z < 0.24$: $\dot{M}_{\rm fb}/\dot{M}_{\rm Edd} \propto M_{\rm BH}^{-0.72}$) and the expected relation $\dot{M}_{\rm fb}/\dot{M}_{\rm Edd} \propto M_{\rm BH}^{-3/2}$. Visual inspection shows that the relation for our sample may be steeper than that found by \citet{Yao2023}.

\subsection{AT2020qhs and the TDE-featureless class} \label{sec:AT2020qhs}
We now turn our attention specifically to AT2020qhs (ID 11), which is a notable event for several reasons. AT2020qhs is a member of the new class of featureless TDEs put forth by \citet{Hammerstein23}. These events are characterized by optical spectra showing a strong blue continuum but with no broad Balmer or \ion{He}{2} emission typical of the optical spectra of TDEs. The peak flare luminosities of these events are several orders of magnitude larger than those of broad-line TDEs, but the rise and fade timescales are similar to the other spectral classes. The host galaxies of TDE-featureless events are typically more massive than broad-line TDEs, suggestive of a higher central black hole mass. Indeed, we find that AT2020qhs possesses the highest black hole mass in our sample, with $\log(M_{\rm BH}/M_\odot) = 8.01 \pm 0.82$. We caution, however, that AT2020qhs is also the highest redshift event in our sample, and as such has the lowest spatial resolution of any event in our sample (4.89 kpc/\arcsec). Additionally, the choice of \msigma~relation can affect the derived black hole mass, which may have implications for the resulting conclusions made here.

\citet{Yao2023} measured the velocity dispersions for two additional TDE-featureless events, AT2020acka \citep{Hammerstein_2020acka, Yao2023} and AT2021ehb \citep{Gezari_2021ehb, Yao2022}, and found corresponding black hole masses of $\log(M_{\rm BH/}/M_\odot) = 8.23 \pm 0.40$ and $\log(M_{\rm BH}/M_\odot) = 7.16 \pm 0.32$, respectively. If we use the \citet{Greene20} $M_{\rm BH} - M_{\rm gal}$ relation for late-type galaxies to estimate the black hole masses for the remaining three featureless events in the \citet{Hammerstein23} sample, AT2018jbv, AT2020riz, and AT2020ysg, we obtain masses within the range $\log(M_{\rm BH}/M_\odot) =$ 6.48 -- 7.70, which are still among the highest masses of those obtained here.

The dependence of the tidal radius and the Schwarzschild radius on the black hole mass is such that above $\sim 10^8 M_\odot$ \citep[sometimes called the ``Hills mass'';][]{Hills75}, a solar-type star will typically pass beyond the black hole's event horizon undisturbed, producing no visible flare. While the black hole mass for AT2020qhs is above this limit, it is still possible to produce an observable TDE around a SMBH of this size. The Hills mass may be exceeded through the disruption of giant stars, although the long timescales and lower luminosities of these events makes it less likely that they will be detected and noted by traditional TDE search methods \citep{Syer1999, MacLeod2012}. This explanation for such a high black hole mass seems unlikely, as the TDE-featureless class is shown to have the highest luminosities of any TDE class while the timescales for these events are comparable to other classes of TDEs \citep{Hammerstein23}.

A more favorable explanation is that the SMBH of AT2020qhs possesses a non-negligible spin which serves to increase the Hills mass \citep{Kesden2012}, as was similarly suggested for the TDE candidate ASASSN-15lh \citep{Leloudas2016}. It has been shown, however, that such SMBHs will contribute only marginally to the overall TDE rate \citep{stone16}. The low predicted rates of spinning SMBHs amongst TDEs may not be a large concern, as \citet{Hammerstein23} noted that most of the TDE-featureless events occur at high redshifts, implying that a larger volume is needed to observe them and hinting at their rarity. Following the work of \citet{Kesden2012} and under the assumption that the disrupted star was of solar type, we can place a lower limit on the spin of the AT2020qhs black hole of $a \gtrsim 0.16$. However, if we instead derive the black hole mass for AT2020qhs using the relation from \citet{xiao11}, the black hole mass becomes $\log(M_{\rm BH/}/M_\odot) = 7.60$, which requires no spin for the disruption of a solar type star.

We note that the disruption of a higher mass star can also potentially explain the black hole mass of AT2020qhs. \citet{Leloudas2015} also addressed this for ASASSN-15lh, finding that only star masses greater than $\sim 3 M_\odot$ can be disrupted by a non-rotating Schwarzschild black hole. These events are also rare \citep{stone16, kochanek16b}, but may be a plausible explanation for AT2020qhs flare. \citet{Mockler22} used measurements of the \ion{N}{3} to \ion{C}{3} ratio in UV spectra to infer the masses of the disrupted stars, finding that the observed ratios necessitate the disruption of more massive stars in the post-starburst hosts they targeted. Larger samples of UV spectra for all TDE types and black hole masses are needed to further investigate whether this is the case for TDE-featureless events such as AT2020qhs.

Spin has been invoked to explain other phenomena observed in TDEs, such as the launching of relativistic jets. Recently, \citet{Andreoni2022} reported the discovery of a jetted TDE in the ZTF survey, concluding that a high spin is likely required to produce such jets. They put a lower limit on the spin parameter of $a \gtrsim 0.3$. \citet{Andreoni2022} also noted the similarities between AT2022cmc and the TDE-featureless class, with the comparable peak flare luminosities and similar lack of broad emission lines in spectra suggesting a connection between the two classes of events. They propose that TDE-featureless events may be jetted TDEs observed off-axis, but further multi-wavelength follow-up of these events is needed to confirm this hypothesis. Nonetheless, the black hole masses AT2020qhs and AT2020acka imply SMBHs with rapid spins and further bolster the possible connection between jetted TDEs and the TDE-featureless class.

\section{Galaxy kinematics and stellar populations} \label{sec:kinematic}
\begin{figure*}
    \centering
    \includegraphics[width=1.5\columnwidth]{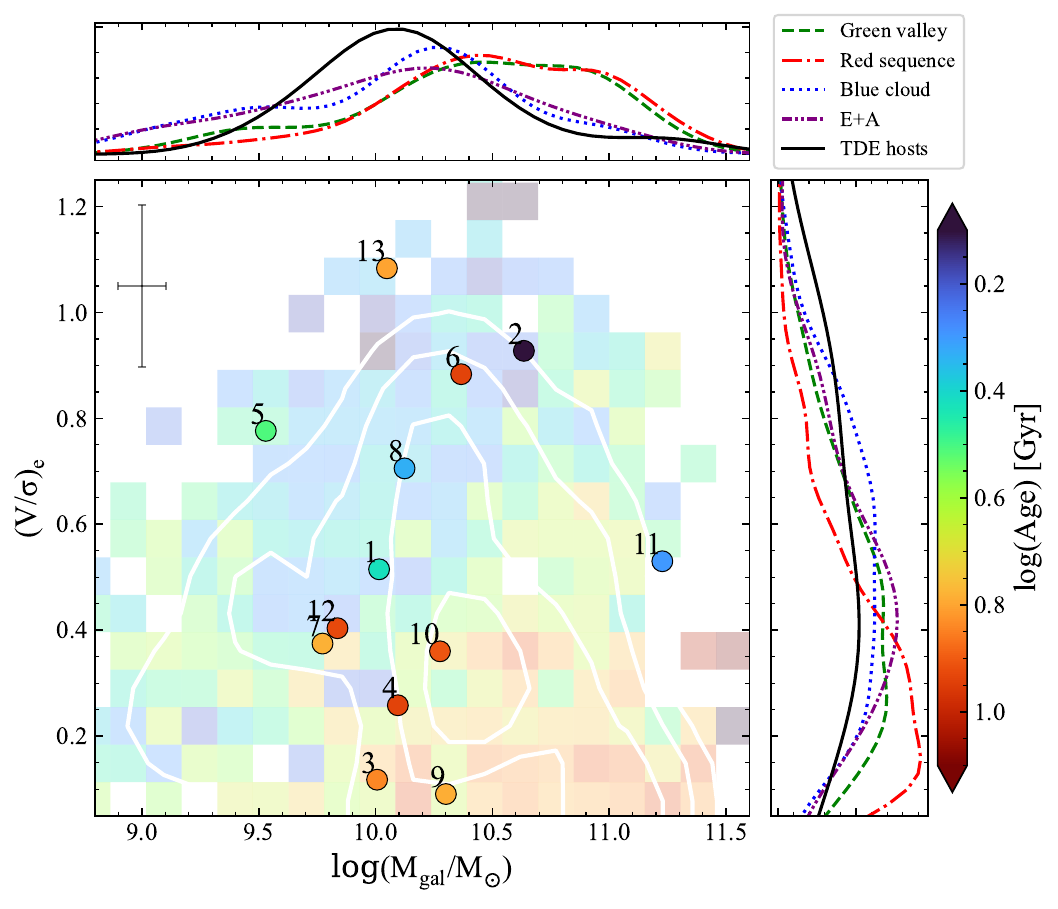}
    \caption{The ratio between stellar ordered rotation and random orbital motion of the TDE host galaxies, defined as $(V/\sigma)_e$, as a function of galaxy stellar mass, with the color of the points/pixels corresponding to the stellar population age. The median uncertainty on the TDE host galaxy values is shown in the top left. Galaxies from the SAMI galaxy survey are shown in the background, with the mean stellar population age of galaxies within a pixel used to determine the pixel color. White contours represent the number density of background galaxies. The top and side panels show the distribution of the TDE hosts and the red sequence, green valley, blue cloud, and E+A galaxies in the background sample obtained by kernel density estimation. We find that the TDE hosts are generally lower mass than most of the background sample, with a larger spread in $(V/\sigma)_e$ than green valley or red sequence galaxies but a distribution similar to E+A galaxies. Labels for each TDE are in Table \ref{tab:sample}.}
    \label{fig:vsigma}
\end{figure*}

We now investigate the kinematic properties on the scale of the effective radius of the entire galaxy light profile ($R_{\rm e, gal}$). Our fits using \ppxf~yield velocities and velocity dispersions, which can be used to estimate the level of rotational support the TDE hosts possess, quantified by the ratio of ordered to random stellar motion $(V/\sigma)_e$, where lower values of $(V/\sigma)_e$ indicate a higher degree of random stellar motions. We adopt the formula of \citet{Cappellari2007}, defined for integral field data:
\begin{equation}
    \left ( \frac{V}{\sigma} \right )^2_e \equiv \frac{\langle V^2\rangle}{\langle \sigma^2\rangle} = \frac{\Sigma_{n=1}^N F_n V_n^2}{\Sigma_{n=1}^N F_n \sigma_n^2},
\end{equation}
where $F_n$ is the flux contained within the $n$th bin, while $V_n$ and $\sigma_n$ are the mean measured velocity and velocity dispersion within that bin. In Figure \ref{fig:vsigma} we show the $(V/\sigma)_e$ for the thirteen TDE host galaxies as a function of stellar population age. We also show the same comparison sample of galaxies as in Figure \ref{fig:ur}. The top and side panels of Figure \ref{fig:vsigma} show the distribution of galaxies in the red sequence, which hosts largely quiescent, elliptical galaxies, the blue cloud, which hosts primarily star-forming galaxies, and the green valley, which hosts recently quenched galaxies, defined from Figure \ref{fig:ur}, E+A galaxies, and the TDE hosts. E+A galaxies from the SAMI survey were selected using the H$\alpha$ equivalent width and Lick H$\delta_{\rm A}$ absorption index using values presented in the MPA+JHU catalogs \citep{Brinchmann04}. We note that only a third of the galaxies in the SAMI survey have a counterpart in the MPA+JHU catalog. The H$\alpha$ equivalent width is limited to $< 4.0$ \AA~and the H$\delta_A$ index is limited to H$\delta_{\rm A} 
 - \sigma$(H$\delta_{\rm A}) > 4.0$ \AA~to isolate post-starburst galaxies.

\citet{vandeSande2018} found a strong correlation between the ratio of ordered rotation to random stellar motion and the stellar population age of a galaxy, such that younger stellar populations are predominantly rotationally supported as in late-type galaxies while older stellar populations are pressure supported by random stellar motions as in early-type galaxies. They also found that $(V/\sigma)_e$ is linked to the observed shape (quantified by the ellipticity $\epsilon$). These correlations link a galaxy's star formation history with its merger history, as mergers will enhance the formation of bulges which in turn lowers a galaxy's $(V/\sigma)_e$ and ellipticity. We find that the TDE host galaxies largely follow this same relation between $(V/\sigma)_e$ and stellar population age, apart from two outliers AT2019azh and AT2020zso (IDs 6 and 13, respectively). AT2019azh is a known E+A galaxy, which have been shown to have varied central stellar population ages and young stellar populations not necessarily confined to the nucleus \citep{Norton2001, Pracy2009}. This may affect the measurement of the host galaxy stellar population age in the central regions in unforeseen ways.

The close link between the merger history, stellar population age, and stellar kinematics is very likely a driving factor behind post-starburst color (used as a proxy for stellar population age) and morphology, and may help explain the TDE preference for such environments. Even before \citet{vandeSande2018} noted the connection between stellar kinematics and stellar population age, \citet{Schawinski10} found that low-mass morphologically early type galaxies in the green valley, which is thought to contain more recently quenched galaxy populations, are linked to mergers which rapidly ushered their migration from the star-forming blue cloud to the green valley and which changed their shape from disk to spheroidal. \citet{Schawinski14} subsequently found that these systems have classic post-starburst populations. However, the majority of galaxies migrate into the green valley through a slow decline in star formation rate likely as a result of gas supply shut off and retain hence their disk shape. The population of green, spiral-like galaxies is noted in \citet{Hammerstein2021}, who compared 19 TDE hosts to red sequence, green valley, and blue cloud galaxies, finding that the TDE hosts are inconsistent with the majority of green valley galaxies which maintained their disk-like morphology inferred through the S\'ersic index.

Given the rate enhancement of TDEs in green valley (and E+A) galaxies, one could expect that TDE host galaxies also cluster in a specific region of $(V/\sigma)_e$. However, we observe a relatively large spread in $(V/\sigma)_e$. The TDE hosts are more evenly distributed in $(V/\sigma)_e$ with a median value of 0.52. We compare the distribution of the TDE host galaxies in $(V/\sigma)_e$ and mass to the red sequence, green valley, and blue cloud galaxies. We find that the TDE hosts, while predominantly green, are generally less massive than the majority of green valley galaxies. This is in agreement with the findings of \citet{Hammerstein2021} for a larger sample of 19 TDE host galaxies from ZTF. The green valley and red sequence distributions in $(V/\sigma)_e$ peak around $\sim 0.2$, indicating that these galaxies are dominated by random stellar motions. In general, we expect a negligible contribution to the TDE rate from stars on circular orbits. This could lead one to conclude that at a fixed stellar mass, a low $(V/\sigma)_e$ might imply a higher TDE rate. However, we should note that the stars within the SMBH sphere of influence (radius $\sim 1$ parsec) contribute only a tiny fraction to the stellar light within the effective radius. Hence the large spread in the $(V/\sigma)_e$ that we observe for the TDE host galaxies cannot directly be translated into a spread in the TDE rate. We thus arrive at the somewhat puzzling observation that the TDE rate appears to be correlated more strongly with the global colors of the host galaxy than the $(V/\sigma)_e$ at its effective radius.

Galaxies that are most certainly dominated by random stellar motions and have stellar populations older than 10 Gyr (i.e., early-type galaxies), have a mean $(V/\sigma)_e = 0.22$. Although three TDE hosts have values around or below this level, they have stellar population ages younger than 10 Gyr at $\sim 7.1$ Gyr. The older, more massive galaxies which are dominated by random stellar motions may also host black holes which exceed the Hills mass, which could explain why the TDE hosts with lower $(V/\sigma)_e = 0.22$ have younger stellar populations than galaxies with similar kinematics. The difference in age between galaxies dominated by random stellar motions and the TDE hosts of similar $(V/\sigma)_e$ implies that the TDE rate likely declines as a galaxy ages despite the increase in the degree of random motion, although the precise reason, whether it be black hole growth beyond the Hills mass or otherwise, and the connection this has with nuclear dynamics is not yet clear given the indirect relationship that these global properties have with factors influencing the TDE rate in the nucleus.

The E+A distribution in $(V/\sigma)_e$ has a mean value of 0.49, similar to the TDE hosts' median value of 0.52. The E+A mass distribution also peaks at $\log(M_{\rm gal}/M_\odot) = 10.07$, while the median TDE host galaxy mass is $\log(M_{\rm gal}/M_\odot) = 10.09$. It is clear that the TDE host galaxies are likely consistent with the same population of galaxies as post-starburst galaxies, which has been suggested previously \citep[e.g.,][]{LawSmith17, Hammerstein2021}. We can also rule out that the TDE hosts come from the same population as red sequence galaxies. An Anderson-Darling test comparing the $(V/\sigma)_e$ of red sequence galaxies to the TDE hosts reveals that the null hypothesis that the two are drawn from the same parent population can be rejected ($p$-value = 0.02). The same cannot be said, however, when comparing green valley galaxies and blue cloud galaxies to the TDE hosts.

The TDE host galaxies also differ in age when compared to the E+A galaxies, with the former having a median stellar population age of 6.12 Gyr, while the E+A galaxies have a mean stellar population age of 2.82 Gyr. One possible conclusion from this is that the TDE host galaxies are post-merger, similar to E+As, but the younger stellar populations produced in the merger-induced starburst having subsided meaning the ages of the stellar populations are older but the other factors which enhance the TDE rate in E+A galaxies (e.g., nuclear star clusters, high central stellar concentrations) remain. Future observations which search for merger signatures, such as in \citet{French2020}, for larger samples of TDEs will be able to confirm the prevalence of post-merger galaxies among TDE host populations. The \galfit~residuals for several galaxies from the LMI data presented here do show remaining features, although differentiating normal dust lane features from true merger signatures like tidal features is difficult. \citet{Stone18} examined factors which enhance TDE rates in post-starburst galaxies, such as SMBH binaries, nuclear stellar overdensities, radial orbit anisotropies, and delay between the initial starburst and the enhancement of the TDE rate due to these factors. This delay time between the initial post-merger starburst and the enhancement of the TDE rate could help to explain why the TDE hosts show older ages but similar global stellar dynamics to the younger post-starburst galaxies. 
  
\section{Conclusions} \label{sec:conclusions}
We have presented the first sample study of IFU observations of thirteen TDE host galaxies from the ZTF survey in order to investigate their kinematic properties and infer their black hole masses. Our main conclusions are as follows:
\begin{itemize}
    \item The black hole mass distribution peaks at $\log(M_{\rm BH}/M_\odot) = 6.05$, consistent with theoretical predictions that TDE populations are dominated by lower mass SMBHs and past observational findings.
    \item There is no significant statistical difference between the X-ray bright and X-ray faint population of TDEs in our sample, which further supports the unifying theory of \citet{Dai2018} that proposes viewing angle effects as the factor which determines X-ray brightness in a TDE.
    \item We find no significant correlation between the black hole masses derived from \msigma~and the black hole masses derived from \texttt{MOSFiT} or \texttt{TDEmass}. This may indicate a need to revisit the way that the black hole mass is imprinted on the light curves of TDEs.
    \item The Eddington ratio is moderately correlated with the black hole mass, although the correlation is likely shallower than the expected relation between the peak fallback accretion rate and the black hole mass, similar to the findings of \citet{Yao2023}. 
    \item We find that the event AT2020qhs, a member of the TDE-featureless class, has the highest black hole mass of the sample: $\log(M_{\rm BH}/M_\odot) = 8.01 \pm 0.82$, above the Hills mass for the disruption of a solar type star. We suggest that the SMBH at the center of this event is rapidly spinning and, assuming that the disrupted star was of solar type, put a lower limit on the spin of $a \gtrsim 0.16$. This further supports the proposed connection between jetted TDEs and the TDE-featureless class put forth by \citet{Andreoni2022}.
    \item We investigate the large-scale kinematics of the TDE host galaxies, particularly the ratio of ordered rotation to random stellar motions $(V/\sigma)_e$, and find that the TDE hosts show similar distributions in $(V/\sigma)_e$ to E+A galaxies but older stellar populations. This may indicate that TDE host galaxies, like E+A galaxies, are post-merger galaxies with the younger stellar populations produced in the merger-induced starburst having subsided, leaving only the older stellar populations. The delay time between post-merger starburst and TDE rate enhancement may also explain the discrepancy in age \citep[e.g.,][]{Stone18}
\end{itemize}

\begin{acknowledgements}
We thank the anonymous referee for their helpful comments towards improving this paper. EH acknowledges support by NASA under award number 80GSFC21M0002.

These results made use of the Lowell Discovery Telescope (LDT) at Lowell Observatory. Lowell is a private, non-profit institution dedicated to astrophysical research and public appreciation of astronomy and operates the LDT in partnership with Boston University, the University of Maryland, the University of Toledo, Northern Arizona University and Yale University. The Large Monolithic Imager was built by Lowell Observatory using funds provided by the National Science Foundation (AST-1005313).

The data presented here were obtained at the W. M. Keck Observatory, which is operated as a scientific partnership among the California Institute of Technology, the University of California and the National Aeronautics and Space Administration. The Observatory was made possible by the generous financial support of the W. M. Keck Foundation. The authors wish to recognize and acknowledge the very significant cultural role and reverence that the summit of Mauna Kea has always had within the indigenous Hawaiian community. We are most fortunate to have the opportunity to conduct observations from this mountain.

The SAMI Galaxy Survey is based on observations made at the Anglo-Australian Telescope. The Sydney-AAO Multi-object Integral field spectrograph (SAMI) was developed jointly by the University of Sydney and the Australian Astronomical Observatory. The SAMI input catalogue is based on data taken from the Sloan Digital Sky Survey, the GAMA Survey and the VST ATLAS Survey. The SAMI Galaxy Survey is supported by the Australian Research Council Centre of Excellence for All Sky Astrophysics in 3 Dimensions (ASTRO 3D), through project number CE170100013, the Australian Research Council Centre of Excellence for All-sky Astrophysics (CAASTRO), through project number CE110001020, and other participating institutions. The SAMI Galaxy Survey website is \hyperlink{http://sami-survey.org/}{http://sami-survey.org}.

The data analysis in this paper was performed on the Yorp and Astra clusters administered by the Center for Theory and Computation, part of the Department of Astronomy at the University of Maryland.
\end{acknowledgements}

\facilities{LDT (LMI), Keck:II (KCWI)}
\software{GALFIT, KCWI-DRP, CWITools, GIST, ppxf}

\bibliography{bibliography, main}
\bibliographystyle{aasjournal}

\end{document}